# Analysis of nonlinear modes of variation for functional data

## Rima Izem and J.S. Marron*

*Department of Statistics, Harvard University*
*Cambridge, Massachusetts, U.S.A.*
*e-mail:* `rizem@fas.harvard.edu`

*Department of Statistics and Operations Research,*
*University of North Carolina at Chapel Hill,*
*Chapel Hill, North Carolina, U.S.A.*
*e-mail:* `marron@email.unc.edu`

**Abstract:** A set of curves or images of similar shape is an increasingly common functional data set collected in the sciences. Principal Component Analysis (PCA) is the most widely used technique to decompose variation in functional data. However, the linear modes of variation found by PCA are not always interpretable by the experimenters. In addition, the modes of variation of interest to the experimenter are not always linear. We present in this paper a new analysis of variance for Functional Data. Our method was motivated by decomposing the variation in the data into predetermined and interpretable directions (i.e. modes) of interest. Since some of these modes could be nonlinear, we develop a new defined ratio of sums of squares which takes into account the curvature of the space of variation. We discuss, in the general case, consistency of our estimates of variation, using mathematical tools from differential geometry and shape statistics. We successfully applied our method to a motivating example of biological data. This decomposition allows biologists to compare the prevalence of different genetic tradeoffs in a population and to quantify the effect of selection on evolution.

Received July 2007.

## Contents

---

*This paper presents some results of the dissertation of the first author under the supervision of the second author.

## 1. Introduction

Researchers in an increasing number of fields, including biology, medicine, and engineering, collect samples of curves, images or objects of common shape. Growth curves of plants and animals, gene expression signals, medical images, and human speech are all real life examples of high dimensional multivariate or functional data, see Dryden and Mardia (1998); Fletcher et al. (2004); Ramsay and Silverman (2005). Understanding the variation in this type of data set is usually of primary interest. Principal Component Analysis (PCA) is the most widely used method to decompose variation in functional data. However, the directions of variation found by PCA are **data driven** and **linear**. So, principal directions are not always easily interpretable by the experimenters and the decomposition fails in the presence of nonlinear variation. The analysis presented in this paper decomposes the variation in terms of modes that are predetermined by the experimenters. Since some of these modes could be nonlinear, we develop a new method, *Template Mode of Variation* or TMV, to decompose variation along nonlinear modes. This decomposition problem is particularly challenging because of the complexity of non-Euclidean geometry. Our method was applied successfully to several sets of reaction norm curves in evolutionary biology, thermal performance curves of caterpillars in Izem (2004) and Izem and Kingsolver (2005), flies in Izem (2004) and viruses in Knies et al. (2006). The matlab code for this decomposition is available at http://www.fas.harvard.edu/~rizem. The method that we present here could be generalized to decompose predetermined modes in any set of curves or images and along directions of interest that satisfy our assumptions.

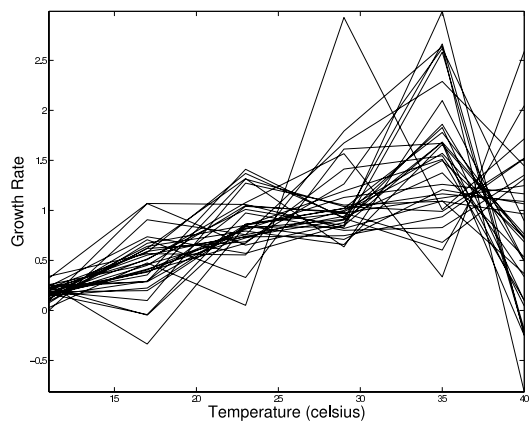

FIG 1. *Thermal Performance Curves of Caterpillars. Each curve represents the family growth rate z at six temperatures t (in Celsius). The between curve variation represents the genetic variation in the population.*

The data and problem which motivated our analysis is presented in Subsection 1.1. We discuss the main assumptions of our analysis in Subsection 1.2. We contrast our approach to other approaches in the literature for analysis of nonlinear variation in Functional Data Analysis (FDA), Shape Analysis, or Manifold Learning in Subsection 1.3. Finally, we summarize the content of the paper in Subsection 1.4.

### 1.1. Motivating data and problem

The data which motivated the analysis in this paper comes from evolutionary biology and is shown in Fig. 1. Each curve shows the growth rate as a function of temperature for a given family of caterpillars, where a family is a set of offsprings of same parents. This data is an example of reaction norm curves, a widely collected data set in biology, where each curve shows the change of a trait as a function of the environment for some genotype, see Scheiner (1993). Thus, in our example, the growth rate is the trait of interest, the temperature is the environmental condition of interest, and each family represents a different genotype in the population. Other examples of reaction norm curves are growth of a virus as a function of temperature in Knies et al. (2006) and photosynthesis of a plant as a function of light intensity or nutrients in the soil.

The curves' variation represents the genetic variation in the population for different environmental conditions. Three gene-environment interactions were identified by biologists as being of particular interest: horizontal shift, vertical shift and generalist-specialist, see Kingsolver et al. (2001). These three variations are shown in Fig. 2. The first mode of variation, the vertical shift, corresponds to a population where some genes clearly dominate other genes in all environments. The two other modes of variation, the horizontal shift and the generalist-specialist, show two different gene-environment trade-offs where all genes are better in some environments and worse in others. In Kingsolver et al. (2001), the given hypothesis is that all these directions exist simultaneously, and there is a need for decomposing the genetic variation in the data into these modes of interest to see which directions are more important.

PCA is one of the most commonly used method to decompose variation in functional data. PCA was applied to the data set in Kingsolver et al. (2004) but it failed to answer the biological questions of interest. In particular, the paper found that the first *linear* principal directions were difficult to interpret biologically and could not distinguish the horizontal shift from the generalist-specialist. PCA failed to give biologically meaningful results in this data set because: first, PCA decomposition is data-driven rather than hypothesis driven; second, PCA can only find linear directions of interest and some of the directions of interest above are nonlinear.

The method described in this paper was successfully applied to decompose the variation in the data shown in Fig. 1 onto the three modes of variation of interest to biologists represented in Fig. 2. Our results are presented in Section 7 and show that the model fits 84% of the variation in the data, most of the variation (73%) is explained by the two nonlinear modes and the remaining 11% is explained by the, linear, vertical shift.

### 1.2. Common shape and predetermined directions

Our analysis assumes that functional data sets have *common shape* and that the modes of variation of interest are predetermined and parameterizable.

In the caterpillar example shown in Fig. 1, we see that each curve has a *common shape.* Each curve increases slowly, tends to reach a maximum and finally decreases rapidly. This common shape assumption is meaningful for several examples in functional data. Because functional data represent different realizations of the same underlying process, growth rate in our example, it is often considered that the variation is around a common template shape reflecting that process, see Wang and Gasser (1997, 1999); Ramsay and Silverman (2005); Dryden and Mardia (1998). In practise, it is often easy to determine the common shape from simply looking at the data. For example: a template gene expression signal could be periodic in time with each period corresponding to a cell cycle, and a template growth curve over time of a plant could be a logistic function. However, the term *common shape* is not easily formally defined. In this paper, the term of common shape or template shape will mean the shape in the *center* of the variation in the data.

There are often some directions of interest identified by the experimenters around this common template shape which we call *modes of variation.* This term was used early on, as in Castro et al. (1986). When these directions are predetermined, the variation in the data is then fully described by the template shape and the modes of variation around the template shape. In our example, we constructed the following 3-parameter Shape Invariant Model (SIM) Lawton et al. (1972) to model the three modes of variation of interest

$$z_i(t_j) = w_i z(w_i(t_j - m_i)) + h_i + \epsilon_{i,j}, \qquad (1.1)$$

where $z_i(t_j)$ is the growth rate of family $i$ at temperature $t_j$. The function $z$ represents the common shape. Parameters $(h_i, m_i, w_i)$ represent the vertical, horizontal and generalist-specialist variation of family $i$ from the template shape. The vertical shift is a linear mode of variation, however since $z$ is not a linear function, the modes of variation parameterized by $m$ and $w$ are nonlinear. Note that the three parameters are on different scales, $h$ is on the growth rate scale, $m$ on the temperature scale and $w$ is unit-less. Thus, it is difficult to compare the magnitude of the variance of different parameters. In addition, it is difficult to quantify the contribution of the variation of each parameter to the between growth curve variation. Our method is a nonlinear ANOVA type decomposition of the variation in growth rate induced by the variation of the parameters. Because our decomposition is done in the same scale as the data, the contributions of the parameters to the total variation in the data are comparable.

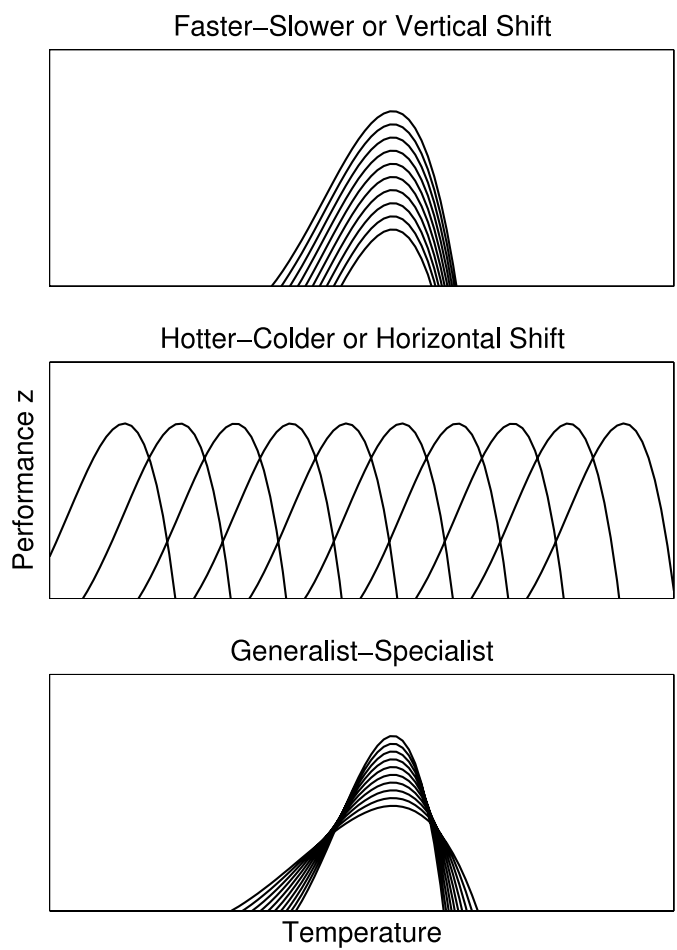

FIG 2. *Three modes of variation of biological interest. Top: Vertical shift of curves:* $z_i(t) = z(t) + h_i$, *where* $z(t)$ *is the common shape curve and* $h_i$ *is the height of family i. Middle: Horizontal shift of curves:* $z_i(t) = z(t - m_i)$, *where* $z(t)$ *is the common shape curve and* $m_i$ *is the location of maximum of family i. Bottom: Generalist-Specialist variation:* $z_i(t) = w_i z(w_i t)$, *where* $z$ *is the common shape curve and* $w_i$ *is the width of the curve of family i.*

The 3-parameter SIM is a particular case of a general Self Modeling Nonlinear Regression (SEMOR) model Kneip and Gasser (1988)

$$z_i(t_j) = R(\theta_i, t_j) + \epsilon_{i,j}, 1 \le i \le n; 1 \le j \le d \qquad (1.2)$$

where the response $z_i$ varies nonlinearly as a function of $t \in \mathbb{R}$, where $\theta_i \in \mathbb{R}^{d'}$ is the vector of parameters of variation, and where $R$ is the common regression function. In the caterpillar example, $\theta_i$ is the vector $(w_i, m_i, h_i)$, and $R(\theta_i, t) = w_i z(w_i(t - m_i)) + h_i$. When $R$ is linear in $\theta_i$ the modes of variation are all linear. Other examples of modes are: variation in frequency and duration of speech signals, or variation in asymptotic height and growth rate of a logistic growth curve of plants. Definitions and results in this paper will be formulated in terms of the general model (1.2). Theoretical results will also assume that the common shape $R$ is known. We discuss how to estimate $R$ from the data using a procrustes type method in Section 6.

### 1.3. Our method and nonlinear variation in the literature

The importance of nonlinear variation of curves around a common shape, such as registration, or horizontal shift, has been recognized for some time in the FDA literature, see Chapter 5 in Ramsay and Silverman (2002, 2005). SIM or SEMOR models are also a common way to account for nonlinearity in FDA, some examples of the extensive literature in the topic are Lawton et al. (1972); Kneip and Gasser (1988); Härdle and Marron (1990); Kneip and Engel (1995); Lindstrom (1995); Ladd and Lindstrom (2000); Brumback and Lindstrom (2004). However, most registration methods which account for nonlinearity of the data have treated nonlinearity as a nuisance. These methods remove nonlinear modes to better estimate the common curve as in Wang and Gasser (1997, 1999), and then to apply conventional linear techniques such as PCA in Silverman (1995). In contrast, motivated by the goals of our collaborators, our premise is that nonlinear variation should be part of the decomposition. We believe that when nonlinear variation is itself of main interest, as is the case in some of the examples cited above, these modes should be properly accounted for in the decomposition. In SIM models, the focus has been on estimating the parameters and the common shape curve in an optimal way, such as in Kneip and Gasser (1988); Härdle and Marron (1990); Kneip and Engel (1995) and quantifying the variability by assuming random parameters and estimating their variance Lindstrom (1995); Brumback and Lindstrom (2004). The variance of each parameter in the SIM model can be used to quantify the variation in the data along a mode. However, as noted in Subsection 1.2, the parameters are on different scales which makes it difficult to compare the contribution of each parameter to the total variation in the data. In addition, the parametrization might not be unique and different parametric scales would have different variance values. In this paper, instead of using the variance of the parameters to quantify variability, we decompose the variation in the natural or intrinsic scale of the data in a nonlinear ANOVA type decomposition. The results of our decomposition are not tied to the parametrization we chose since an equivalent parametrization would produce the same space of variation. Thus, an equivalent parametrization of the modes of variation will produce the same decomposition. Our decomposition is also invariant to a linear transformation of the data, since a linear transformation of the data will be absorbed by the common shape function.

In Shape Analysis, the notion of mean and variance in a metric manifold were defined in Fréchet (1948) as the Fréchet mean and Fréchet variance. These notions are widely studied in the field of robust statistics and have more recently been used by probabilists for characterizing distributions of shape data, see Kume and Le (2000); Le (2001); Bhattacharya and Patrangenaru (2003, 2005). These authors showed fundamental results for the Fréchet mean, conditions of its existence, consistency of its estimates from the sample, and a central limit theorem for the Fréchet mean on a Riemannian manifold. Our focus in this paper is not in estimating the Fréchet mean, but in decomposing variability. One drawback of the Fréchet variance is that it is a number regardless of the dimensionality of the manifold. It represents the total variation in the data, but it does not offer the full understanding of the variation that the variance-covariance matrix offers in the Euclidean case. In our approach, we decompose the Fréchet variance $\widetilde{SSM}$ into meaningful quantities, each representing variation along a mode of interest. i.e.

$$\widetilde{SSM} = \sum_{i=1}^{d'} \widetilde{SSM}_i$$

where $\widetilde{SSM}_i$ quantifies the variation along mode $i$. This goal is reached by defining new metrics which allow for the decomposition of the variation. More precisely, we define a new $d_V$ in the manifold such that,

$$\widetilde{SSM} = \sum_{i=1}^{n} d_V^2(R(\theta_i, \mathbf{t}), \tilde{R})$$

where $\tilde{R}$ is a Fréchet mean. In the one dimensional case, the metric is simply the *arcdistance* or shortest distance between two points along the curved one dimensional differentiable manifold. In the two dimensional case we use a Pythagorean like formula to define a path metric in the manifold as a function of the arcdistances along each mode. Contrary to the Euclidean case, the path metric between two points along one-dimensional geodesics are not unique. So, our final distance accounts for this multiplicity by assigning weights for each path metric.

Manifold learning methods as in Hastie and Stuetzle (1989); Tenenbaum et al. (2000); Donoho and Grimes (2003); de Silva and Tenenbaum (2003a); de Silva and Tenenbaum (2003b); Fletcher et al. (2004) extend the dimensionality reduction aspect of PCA to multivariate data lying in a nonlinear manifold. They are data-driven and exploratory, so the results they find are not always interpretable. Moreover, these methods do not attempt to quantify the variation along given nonlinear principal directions. In contrast, in our approach, we decompose the variation into pre-identified, and interpretable, modes given by the experimenter. Lastly, our decomposition exploits the fact that the data of interest to us are not arbitrary multivariate data but instead are functional data, with approximately common shape.

### *1.4. Structure of this paper*

We illustrate the geometry of nonlinear spaces of variation with four toy examples in Section 2. Two toy examples illustrate one dimensional spaces of variation, and two other toy examples illustrate two dimensional spaces of variation. In Section 3 we define the Fréchet mean and variance for functional random variables varying in a manifold. We also propose a ratio measuring the variation along nonlinear modes. Section 4 discusses our choice of metrics in the manifold and the proposed quantification of one, two or multiple simultaneous modes of variation. In the one-dimensional case, the metric is simply the arcdistance. In the two-dimensional case and higher, the choice of metric is key to decomposing variability. The metric in higher dimensions is defined by a Pythagorean like formula using all the one dimensional arcdistances. The consistency of our estimates is shown in Section 5. We discuss the implementation of this method as well as present results of our method on the motivating example in Section 6. We finally discuss the result of the decomposition to the motivating data set in Section 7.

### 2. Geometry of the space of variation

Before considering the space of variation in the most general case in model (1.2), we explore four particular examples with a given common shape. The first two examples are one mode cases, the two other examples are two-mode cases. The one mode cases illustrate what the geometry of the space of variation is for a linear and a nonlinear mode, they are:

(a) Linear mode example: vertical shift of curves around a common shape. In this example, $\theta_i = h_i$, and $R(\theta_i, t) = z(t) + h_i$ in model (1.2).

(b) Nonlinear mode example: horizontal shift of curves around a common shape. In this example, $\theta_i = m_i$, and $R(\theta_i, t) = z(t - m_i)$ in model (1.2).

The two modes examples are:

(c) Linearly separable model example: simultaneous variation of curves along the vertical shift and horizontal shift. In this example $\theta_i = (h_i, m_i)$, and $R(\theta_i, t) = z(t - m_i) + h_i$ in model (1.2). We call this model linearly separable because we can write the function $R((m_i, h_i), \mathbf{t})$ as the sum of two different functions, each depending on only one parameter. More precisely, $R((m_i, h_i), t) = R_1(m_i, t) + R_2(h_i, t)$ where $R_1(m_i, t) = z(t - m_i)$ and $R_2(h_i, t) = h_i$. As we will see in Section 4, the geometry of the space of variation and the decomposition will be simpler to describe in this case because of this property.

(d) General case example: simultaneous variation of curves along the horizontal shift and generalist-specialist. In this example, $\theta_i = (m_i, w_i)$, and $R(\theta_i, t) = w_i z(w_i(t - m_i))$ in model (1.2). As we will see in Section 4 defining a distance that allows the decomposition will be less trivial in this case, but will rely on similar ideas defined for one-mode and two-mode linearly separable examples.

For illustration, the curve variation as well as the space of variation is shown for each of these four examples with a parabola common shape in Fig. 3 and Fig. 4. These four examples and their illustrations are discussed in details in Subsection 2.1. In Subsection 2.2, we present the general result on the dimensionality and geometry of the space of variation for a set of curves of common shape.

### 2.1. Examples in one and two dimensions

To visualize the variation in the curve space and in the point cloud space, we sample $z$ at three distinct points $(t_1, t_2, t_3)$. As illustrated in Figs. 3 and 4, sampling at three points allows for a representation of an infinite dimensional curve as a point in three dimensions. So, a set of parabolic curves in the curve space appears as a set of points in the point cloud space and the variation in the curves corresponds to variation of the points.

The linear mode example, the vertical shift, is presented in the two left panels of Fig. 3. We see that when there is one mode of variation of the curves and it is linear, here vertical shift, the points in the point cloud space fall along a line. We call the line in this example the space of variation of the data. Note that when the mode of variation is linear, the arithmetic average falls within the space of variation. Note also that deviation of each data point from the mean could be measured by the Euclidean metric. The nonlinear mode example is presented in the two right panels of Fig. 3. We see that when there is one mode of variation of the curves and it is nonlinear, here the horizontal shift, the points in the point cloud space fall along a curve. We call this curve in this example the space of variation of the data. Note that when the mode of variation is nonlinear, the arithmetic mean will not fall in the space of variation.

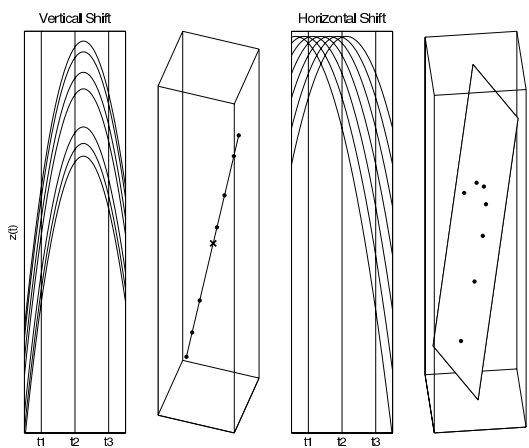

FIG 3. *Illustration of the vertical shift and horizontal shift modes in the curve space and the point cloud space. Two examples of one dimensional mode of variation, one linear and one nonlinear. From left to right, Panel 1: Parabolas vertically shifted. Panel 2: Visualization of the variation in 3d. Since the mode of variation is linear, the space of variation is a line and the mean (x) lies in the space of variation. Panel 3: Parabolas horizontally shifted. Panel 4: Visualization of the variation in 3d. Since the mode of variation is not linear, the space of variation is a 1-dim manifold or curve.*

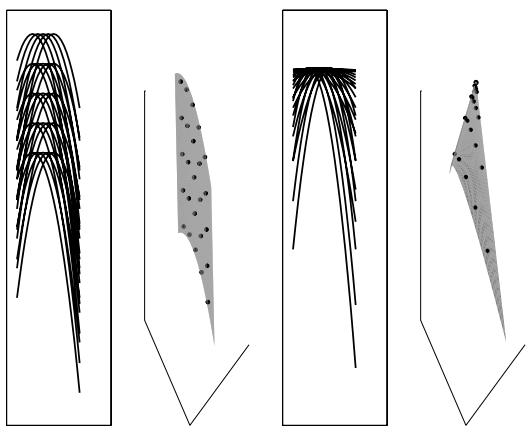

FIG 4. *Curve space, and space of variation in the two dimensional case. From left to right, Panel 1: Toy parabola curves varying simultaneously along the vertical shift (parameterized by h) and the horizontal shift (parameterized by m). Panel 2: The space of variation corresponding to Panel 1. It is a surface. Panel 3: Toy parabola curves varying simultaneously along the generalist-specialist (parameterized by w) and horizontal shift (parameterized by m). Panel 4: The space of variation corresponding to Panel 3. It is a surface. Each curve in the curve spaces in Panel 1 and Panel 3 corresponds to a point in the space of variation in Panel 2 and Panel 4.*

Similarly, using the Euclidean distance will not be the best characterization of deviation along the space of variation. For the one mode case, we will propose in Section 3 and Section 4 to use the arcdistance, or distance along the manifold, instead of the Euclidean distance as measure of proximity. We also propose to use the Fréchet mean and variance, or mean and variance along the manifold, as a measure of center and spread in the manifold.

Fig. 4 illustrates the case of simultaneous variation along two modes, vertical shift and horizontal shift in the left two panels, horizontal shift and generalist-specialist in the right two panels. We see in both examples that when there are two modes of variation, the points in the point cloud space fall in a plane or a surface. We call the surface in each example the space of variation of the data. When the space of variation is curved, the mean does not fall in the space of variation and the Euclidean metric is not appropriate for proximity. As in the one-dimensional case, we propose in Section 3 to use the Fréchet mean set for a center in the manifold of variation. However, in contrast to the one-dimensional case, we do not use the arcdistance or shortest distance along the manifold as measure of proximity. Instead, we define in Section 4 new metrics by a Pythagorean like formula using the one-dimensional arcdistances along each mode. The advantage of these new metrics is that, by construction, we can decompose variation in the manifold in an ANOVA type decomposition along the modes of interest. As discussed in Section 4, the construction of this metric is simpler in the linearly separable case than in the general case.

## 2.2. General dimension

For a variation or a set of variation around a common shape, the space supporting the points in the point cloud space is called the space of variation. In the one-mode examples, the line or the curve were the space of variation. In the two-modes examples, the surface is the space of variation. In general, given data as in model (1.2) with a common shape $R(\theta, t)$ and parameter space $\Theta \subset \mathbb{R}^{d'}$, we define the *space of variation* $V$ as the subspace of $\mathbb{R}^d$ such that,

$$V = \{(x_1, \ldots, x_d) \in \mathbb{R}^d; \exists \theta \in \Theta \text{ such that } x_j = R(\theta, t_j) \forall j = 1, \ldots, d.\}$$

Theorem 2.1 states that under certain conditions on the common shape and the parameter space, the space of variation is a manifold of the same dimension $d'$ as the parameter space

**Theorem 2.1.** *For a fixed sampling vector* $t \in \mathbb{R}^d$, *let the function* $R(., \mathbf{t})$ *be such that*

$$R(., \mathbf{t}) : \mathbb{R}^{d'} \quad \rightarrow \quad V$$
$$\theta \quad \mapsto \quad R(\theta, \mathbf{t})$$

*If* $R(., \mathbf{t})$ *is an homeomorphism from* $\mathbb{R}^{d'}$ *to* $V$, *then* $V$ *is a manifold of dimension* $d'$.

See proof in Appendix. For the caterpillar data, this condition is satisfied for any polynomial common shape $z$ of degree higher than 2 and the three modes of variation parameterized by $(w, m, h)$ in model (1). The metrics $d_V$ which we define in the manifold of variation $V$ in Section 4 using the arcdistances along each mode will additionally require that $R(., \mathbf{t})$ be differentiable.

## 3. Variation in a manifold

In the previous section, we visualized the space of variation in the case of a one-dimensional mode or a two-dimensional simultaneous modes. The geometry of the space is non-euclidean in the presence of nonlinear variation, so the usual notions of mean and variance are not representative of the center and spread of the distribution. We first define in this section an appropriate mean, as a measure of center of variation, and variance, as a measure of spread along the manifold. The goal is to use these new mean and variance to define a ratio to quantify nonlinear modes.

### 3.1. Fréchet mean and Fréchet variance

It is well known that given a set of data in a nonlinear manifold, the arithmetic mean will not necessarily fall in the manifold. Similarly using the Euclidean distance as a measure of variation around the mean shape is not appropriate in the manifold because this distance is not a good measure of proximity in a nonlinear space. These facts have been discussed extensively in the shape analysis literature, and illustrated with a toy example in functional data in Izem et al. (2003). Because the usual definitions of mean and variance in a linear space are not meaningful measures of center and spread in a nonlinear space, Fréchet generalized the notions of mean and variance to manifolds in Fréchet (1948). The generalization proceeds as follows, for a real random variables $X$ in a Euclidean space with measure $\mu$, the expected value $E(X)$ is the point in the space which minimizes the variance, i.e. let $F(y) = \int ||X - y||^2 d\mu(x)$, then

$$E(X) = \operatorname{argmin}_{y \in \mathbb{R}}(F(y)) \text{ and } Var(X) = F(E(X)).$$

For a metric manifold $(M, d)$ with measure $\mu$, let the Fréchet function be,

$$F(y) = \int_M d^2(x, y) d\mu(x).$$

This function is well defined if $\int_M d^2(x, y) d\mu(x) < \infty, \forall y \in M$

In a metric manifold $(M, d)$, the Fréchet mean set $E_F(X)$ of a random variable $X$, in the manifold $M$, with probability measure $\mu$, is the set of points on the manifold which minimize the function $F(y)$. The Fréchet variance $\widetilde{Var}_F(X)$ is the value $F(\tilde{X}_F)$ for any $\tilde{X}_F$ in the Fréchet mean set $E_F(X)$. i.e.

$$\tilde{X}_F \in E_F(X) \quad \text{iff} \quad F(\tilde{X}_F) = \inf_{y \in M} F(y).$$
$$\widetilde{Var}_F(X) \quad = \quad F(\tilde{X}_F)$$

Note that although we can have more than one Fréchet mean, we can define a unique Fréchet variance. Note also that the Fréchet variance is a scalar, and not a matrix, even for manifolds of dimension $d' \geq 2$. In this paper, we are concerned with functional data $X$, so the Fréchet mean set is a set of functional data.

### 3.2. *Quantifying the variation*

In our context, let $z_1, \ldots, z_n$ be a functional data set of common shape as in model (1.2), where each $z_i$ is associated to the regression functions $R_i$ and parameter $\theta_i$ in a closed set $\Theta$. Thus, $R_i = R(\theta_i, \mathbf{t})$ is in the metric manifold $(V, d_V)$ and an intuitive estimate of the Fréchet function is

$$F_n(R) = \sum_{i=1}^{n} d_V^2(R_i, R).$$

We can also derive an estimate of the Fréchet mean set $E_{F,n}(R)$ and an estimate of the Fréchet variance $\frac{1}{n}\widehat{SSM}$, where

$$\tilde{R} \in E_{F,n}(R) \quad \Leftrightarrow \quad F_n(\tilde{R}) = \min_{R \in V} F_n(R)$$

$$\widehat{SSM} = \sum_{i=1}^{n} d_V^2(R_i, \tilde{R})$$

We propose to quantify the variation in the manifold by the following ratio,

$$\widehat{RSS} = \frac{\widehat{SSM}}{\widehat{SSM} + SSE} \tag{3.1}$$

where

$$SSE = \sum_{i=1}^{n} ||z_i - R_i||^2$$

Note that for linear modes, if we take $d_V$ to be the Euclidean metric, this ratio corresponds to the usual ratio of sums of squares used in PCA. Note also that the choice of the distance $d_V$ is critical, it is the main building block to the decomposition we show in this paper. In one dimension, we choose $d_V$ to be the arcdistance, or the distance along the curved one dimensional manifold. For a differentiable one-dimensional manifold with parametrization $\gamma : \theta \in \mathbb{R} \mapsto \gamma(\theta) \in \mathbb{R}^d$, we have that

$$\mathrm{Arcd}(\gamma(\theta_1), \gamma(\theta_2)) = \int_{\theta_1}^{\theta_2} ||\frac{\partial}{\partial \theta}\gamma(\theta)|| d\theta; \forall \theta_1, \theta_2. \tag{3.2}$$

In higher dimensional spaces of variation, our choice of $d_V$ is motivated by our decomposition goal, i.e. we define $d_V$ such that

$$\text{by definition,} \qquad \widetilde{SSM} = \sum_{i=1}^{n} d_V^2(R_i, R), \qquad (3.3)$$

$$\text{by construction} \qquad \sum_{i=1}^{n} d_V^2(R_i, R) = \sum_{k=1}^{d'} \widetilde{SSM}_k \qquad (3.4)$$

$$\text{and} \qquad \widetilde{RSS}_k = \frac{\widetilde{SSM}_k}{\widetilde{SSM} + SSE} \qquad (3.5)$$

where $\widetilde{RSS}_k$ quantifies the variation along mode $k$. Constructing the distance $d_V$ is challenging in manifolds since the notion of orthogonality is local whereas it is global in linear spaces. We reach this goal by incorporating in $d_V$ the arcdistances along each mode in a Pythagorean like formula. We first define these distances more formally in Section 4. We show in Section 5 that in each case, with our choice of distance, the estimates above are consistent estimates of the Fréchet mean and variance. Moreover, we find that in the one dimensional and linearly separable cases the Fréchet mean and its estimate (for a given $n$) are unique. Finally, we show in the general case, consistency. Using these estimates of Fréchet mean and variance, we use the proposed the above ratios $\widetilde{RSS}_k$'s to quantify nonlinear modes in a manifold.

## 4. Distance and decomposition of variation in the manifold

In the one dimensional case, the metric $d_V$ is simply the arcdistance and the quantification is easy to do as seen in Subsection 4.1. In the two-dimensional case and higher, the geodesic distance does not allow for a straightforward decomposition. So we define the metric as a function of the arcdistances along each mode. As seen in Subsection 4.2, the expression of this metric is easy in the linearly separable case. The distance is not trivial to generalize to the general, non linearly separable, case. We first see how to generalize it in the two-dimensional case in Subsection 4.3. Finally, Subsection 4.4 discusses the generalization of this metric to higher dimensional spaces of variation.

### *4.1. One mode*

A natural choice of distance $d_V(x, y)$ between two points $x$ and $y$ in the one-dimensional manifold of variation is the arcdistance, $Arcd(x, y)$. We show in Section 5 that in the one dimensional case, the ratio $\widetilde{RSS}$ defined in Equation (3.1) is well defined, the Fréchet mean is unique, its estimate for a sample of points of size $n$ is also unique, and this estimate is consistent.

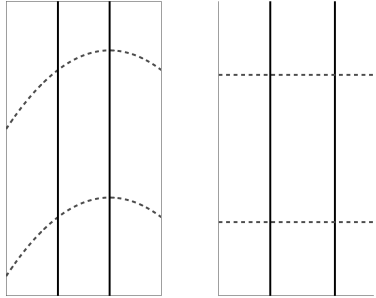

FIG 5. *Example of a space satisfying the equality of path condition and a representation of the transformation of the space into $\mathbb{R}^2$. Left: Representation of the space of variation when the two modes are vertical shift (parameterized by h) and horizontal shift (parameterized by m). Right: Representation of the transformation of the space of variation. The distances along the curves (dashed line type) and the distance along the lines (solid line type) are the same as in the original space of variation in the left panel.*

### 4.2. Multiple simultaneous modes, linearly separable case

We described in Section 2 an example of a 2-dimensional linearly separable space of variation. It is shown in the two left panels of Fig. 4, the two modes of variation are the vertical shift and the horizontal shift. We first review the equality of path property in Subsection 4.2.1 and define it in the general case. In Subsection 4.2.2, we use the property to construct the metric and the decomposition of variation.

#### 4.2.1. Linear separability and equality of path

We illustrate the geometry of the space of variation in the two dimensional, linearly separable case, in Fig. 5. We see in the left panel of Fig. 5 four points $\{R((m_i, h_j), \mathbf{t})\}_{i,j=1,2}, m_1 < m_2; h_1 < h_2\}$ in the space of variation. The two parallel solid lines correspond to the vertical shift variation for two different fixed values of location parameters $m_1$ and $m_2$. Similarly, the two parallel dashed curves correspond to the horizontal shift variation for two different fixed values of the height parameter $h_1$ and $h_2$. To go from one point $R((m_1, h_1), \mathbf{t})$ to another $R((m_2, h_2), \mathbf{t})$, by two steps, along the curves of variation, we have two possible paths. The first path goes first along the top dashed curve $m \in (m_1, m_2) \mapsto R((m, h_1), \mathbf{t})$, then along the left solid line $h \in (h_1, h_2) \mapsto R((m_2, h), \mathbf{t})$. The second path goes first along right solid line $h \in (h_1, h_2) \mapsto R((m_1, h), \mathbf{t})$, then along the bottom dashed curve $m \in (m_1, m_2) \mapsto R((m, h_2), \mathbf{t})$. For this special case, the two solid line segments and the two dashed curve segments from the two different paths are of equal length. This *equality of path property* is satisfied by any manifold generated by a linearly separable model. Namely, the arcdistance between two points along the space of variation if we fix one of the parameters, depends only on the other parameter. For our example, we have that

$$|R(m, h_1) - R(m, h_2)|^2 = (z(t-m) + h_1 - z(t-m) - h_2)^2 = (h_1 - h_2)^2; \forall h_1, h_2, m$$

so this distance along the surface of variation between the points $(m, h_1)$ and $(m, h_2)$ depends only on $h_1$ and $h_2$ and not on $m$. In addition, we have as in Equation 3.2 that

$$\text{Arcd}(R(m_1, h), R(m_2, h)) = \int_{m_1}^{m_2} ||\frac{\partial}{\partial m} z(\mathbf{t} - m)||dm; \forall m_1, m_2, h$$

Since the derivative $\frac{\partial}{\partial m} z$ does not depend on $h$, the distance along the surface of variation between the points $(m_1, h)$ and $(m_2, h)$ depends only on $m_1$ and $m_2$ and not on $h$. In general, we define a linearly separable model as follows.

**Definition 4.1.** *A regression function $R(\theta, t)$ in model (1.2) for parameter $\theta = (\theta_1, \ldots, \theta_{d'})$ in $\mathbb{R}^{d'}$ is linearly separable if there exists $d'$ differentiable functions $R_1, \ldots, R_{d'}$ such that*

$$R(\theta, t) = \sum_{i=1}^{d'} R_i(\theta_i, t)$$

The linearly separable model satisfies the property of equality of paths. We say that we have equality of paths in a $d'$ dimensional manifold of variation $V$ with parameter space $\Theta$ if the distance between two points, which share the same values in $d' - 1$ parameters and differ on one parameter, depend only on the varying parameter. i.e. for all $(\theta_{j,1})_{1 \le j \le d'}, (\theta_{j,2})_{1 \le j \le d'}$ in $\Theta$ and for all $j$

$$\text{Arcd}\left(R(\theta_{1,1}, \ldots, \theta_{j,1}, \ldots, \theta_{d',1}; \mathbf{t}), R(\theta_{1,1}, \ldots, \theta_{j,2}, \ldots, \theta_{d',1}; \mathbf{t})\right) = C_{\theta_{j,1}, \theta_{j,2}},$$

where $C_{\theta_{j,1}, \theta_{j,2}}$ is a function depending only on the values of $\theta_{j,1}$ and $\theta_{j,2}$ of the $j$th parameter, and not on the values of other parameters $(\theta_{i,1})_{1 \le i \le d'; i \ne j}$.

### 4.2.2. Distance and decomposition

Using this property, it is straightforward to define a metric in the manifold which will decompose the total variation in the data onto the modes of interest using the following metric

**Theorem 4.1.** *For a $d'$ $(d' \ge 2)$ differentiable manifold of variation $V$ generated by a linearly separable model the non-negative function $d_V$ defined below is a distance in $V$*

$$d_V : V \times V \quad \rightarrow \quad \mathbb{R}^+$$

$$(R((\theta_{1,1}, \ldots, \theta_{d',1}), \mathbf{t}), R((\theta_{1,2}, \ldots, \theta_{d',2}), \mathbf{t})) \quad \mapsto \quad \sqrt{\sum_{i=1}^{d'} C^2_{\theta_{i,1}, \theta_{i,2}}}$$

We show that $d_V$ satisfies the conditions of a distance in the Appendix. The proof shows that the space of variation for a linearly separable model is homeomorphic to a subspace of $\mathbb{R}^{d'}$. Furthermore, the function $d_V$ we defined above corresponds to the Euclidean metric in this subspace of $\mathbb{R}^{d'}$. The right panel of Fig. 5 illustrates this mapping in the case of $d' = 2$. This mapping preserves the arcdistances along the curves of the modes of variation, and $d_V$ corresponds to the Euclidean distance in this mapping.

Finally, we have a simple decomposition in the linearly separable case. Recall that by definition $\widetilde{SSM} = \sum_{i=1}^{n} d_V^2(R_i, \tilde{R})$. By construction from the above theorem, we have that

$$d_V^2(R_i, \tilde{R}) = \sum_{k=1}^{d'} C_{i,k}^2$$

where $C_{i,k}^2$ is a function of the $k$th mode. So, finally we have the decomposition

$$\widetilde{SSM} = \sum_{k=1}^{d'} \widetilde{SSM}_k, \text{ where } \widetilde{SSM}_k = \sum_{i=1}^{n} C_{i,k}^2 \text{ and } \widetilde{RSS}_k = \frac{\widetilde{SSM}_k}{\widetilde{SSM} + SSE}$$

By construction, $\widetilde{RSS}_k$ quantifies the variability around the Fréchet mean along mode $k$.

### *4.3. Two modes, general case*

We described in Section 2 an example of a 2-dimensional general case space of variation. We saw in the two right panels of Fig. 4 that when the two modes of variation are horizontal shift and generalist-specialist, the space of variation is a two-dimensional manifold. The right panel of Fig. 6 is a representation of the same space. In this figure, the three parabolic curves correspond to the horizontal shift variation for three different fixed values of the width parameter and the three lines correspond to portions of the curve representing the generalist-specialist mode for three different fixed values of the horizontal shift variation. As illustrated in Fig. 6, in the general case we do not have the equality of path property. More precisely, the paths for going in two steps from one point to another in the manifold will not be equal. We define in Subsection 4.3.1 and Subsection 4.3.2 two different metrics $d_{1,O}$ and $d_{2,O}$ by considering two different mappings $L_{1,O}$ and $L_{2,O}$ of the space of variation into a subspace of $\mathbb{R}^2$. Each metric is homeomorphic to the Euclidean distance in the transformed subspace of $\mathbb{R}^2$. We finally combine the two distances in Subsection 4.3.3 to define a metric $d_V$ in the manifold indexed by the weight $\gamma$ as

$$d_V(x,y) = \sqrt{\gamma d_{1,O}(x,y)^2 + (1-\gamma)d_{2,O}(x,y)^2}.$$

As discussed in Remark 4.1, we chose $\gamma$ to be $1/2$ in our analysis. We see in Subsection 4.3.4 how we can use this distance for the decomposition of interest.

Because each mapping is a different transformation of the manifold into a plane, we need to define each transformation with respect to an origin $O = R_{0,0} = R((\alpha_0, \beta_0), \mathbf{t})$ in the space. We defer the discussion on the choice of origin to Section 5.

### 4.3.1. First transformation

The first transformation $L_{1,O}$ maps the space of variation $V$ into $\mathbb{R}^2$ by preserving the arcdistances along the two curves which cross the origin $O$. The two curves which cross at the origin are the dotted curves in Fig. 6, and the image of the transformation $L_{1,O}$ is shown in the bottom right panel of Fig. 6. We define the distance $d_{1,O}$ in $V$ as the Euclidean distance in $L_{1,O}(V)$. More formally, let $R_{i,k} = R((\alpha_i, \beta_k), \mathbf{t})$ for $i, k = 0, 1, 2$, then

$$\begin{aligned}
L_{1,O} : V &\rightarrow \mathbb{R}^2 \\
R_{1,1} &\mapsto (\eta, \zeta) \text{ such that} \\
\eta &= \text{sign}(\alpha_1 - \alpha_0)\text{Arcd}(R_{1,0}, R_{0,0}) \\
\zeta &= \text{sign}(\beta_1 - \beta_0)\text{Arcd}(R_{0,1}, R_{0,0})
\end{aligned}$$

After transformation, we define the metric $d_{1,O}$ between two points in the manifold as equivalent to the Euclidean distance in the space $L_{1,O}(V)$ such that

$$\begin{aligned}
d_{1,O} : V \times V &\rightarrow \mathbb{R}^+ \\
(R_{1,1}, R_{2,2}) &\mapsto d_{1,O}(R_{1,1}, R_{2,2}) \\
d_{1,O}(R_{1,1}, R_{2,2}) &= ||L_{1,O}(R_{1,1}) - L_{1,O}(R_{2,2})||
\end{aligned}$$

The following Proposition states that this function defines a metric,

**Proposition 4.1.** *For all origins $O$, and for a differentiable manifold $V$, the function $d_{1,O}$ defines a metric on $V$.*

This Proposition is equivalent to showing that $L_{1,O}$ is isometric. By using simple algebra, we can further simplify the expression of $d_{1,O}$ as a function of the parameters of variation rather than the linearizing function as,

**Corollary 4.1.**

$$d_{1,O}^2(R_{1,1}, R_{2,2}) = Arcd^2(R_{1,0}, R_{2,0}) + Arcd^2(R_{0,1}, R_{0,2})$$

### 4.3.2. Second transformation

The second transformation $L_{2,O}$ maps $V$ into $\mathbb{R}^2$ by preserving the arcdistances along the curves which cross the points of interest. For example, the curves which cross at point 1 in Fig. 6 are the two dashed curves. Similarly, the two curves which cross at point 2 in Fig. 6 are the two dotted and dashed curves. The image of the space of variation by this transformation $L_{2,O}$ is shown in the top right panel of Fig. 6. As for the first transformation, we define a metric in $V$ as the Euclidean distance in $L_{2,O}(V)$. More formally, the second transformation $L_{2,O}$ is defined as follows,

$$
\begin{aligned}
L_{2,O} : V &\rightarrow \mathbb{R}^2 \\
R_{1,1} &\mapsto (\eta, \zeta) \text{ such that} \\
\eta &= \text{sign}(\alpha_1 - \alpha_0)\text{Arcd}(R_{1,1}, R_{0,1}) \\
\zeta &= \text{sign}(\beta_1 - \beta_0)\text{Arcd}(R_{1,1}, R_{1,0})
\end{aligned}
$$

We can then define a function $d_{2,O}$ which corresponds to the Euclidean distance in the linearized space.

$$
\begin{aligned}
d_{2,O} : V \times V &\rightarrow \mathbb{R}^+ \\
d_{2,O}(R_{1,1}, R_{2,2}) &= ||L_{2,O}(R_{1,1}) - L_{2,O}(R_{2,2})||
\end{aligned}
$$

We can rewrite $d_{2,O}$ as a function of the parametrization as follows

$$
\begin{aligned}
d_{2,O}(R_{1,1}, R_{2,2}) &= \sqrt{A^2 + B^2}, \text{ where} \\
A &= \text{sign}(\alpha_1 - \alpha_0)\text{Arcd}(R_{1,1}, R_{0,1}) \\
&\quad - \text{sign}(\alpha_2 - \alpha_0)\text{Arcd}(R_{2,2}, R_{0,2}) \\
B &= \text{sign}(\beta_1 - \beta_0)\text{Arcd}(R_{1,1}, R_{1,0}) \\
&\quad - \text{sign}(\beta_2 - \beta_0)\text{Arcd}(R_{2,2}, R_{2,0})
\end{aligned}
$$

**Proposition 4.2.** *The function $d_{2,O}$ satisfies the following conditions: (i) $d_{2,O}$ is non-negative, (ii) $d_{2,O}$ is symmetric. (iii) $d_{2,O}$ satisfies the triangular inequality. Moreover, if the mapping $L_{2,O}$ is one-to-one, then $d_{2,O}$ satisfies the condition*

$$d_{2,O}(R_{1,1}, R_{2,2}) = 0 \Leftrightarrow R_{1,1} = R_{2,2}$$

*which, in addition to the above conditions, makes $d_{2,O}$ a distance.*

Note that if the space $V$ is linearly separable, then the two functions $d_{1,O}$ and $d_{2,O}$ are the same, and they are equal to the distance defined in the linearly separable case. This result is stated in the following proposition and proved in the Appendix

**Proposition 4.3.** *When $V$ is linearly separable, $d_V$ is the distance defined in Subsection 4.2.2, and $d_{1,O}$ and $d_{2,O}$ are as defined above, we have that $d_{1,O} = d_{2,O} = d_V$ for all origins $O$ in the manifold.*

### 4.3.3. Distance in the manifold

The fact that $d_{2,O}$ is not always a distance in the general case is not an issue since we can combine it with $d_{1,O}$ to define a distance in the space of variation as stated in Proposition 4.4 proved in the Appendix.

**Proposition 4.4.** *The non-negative function $d_{V,O,\gamma}$ defined as*

$$
\begin{aligned}
d_{V,O,\gamma} : V \times V &\rightarrow \mathbb{R}^+ \\
(R_{1,1}, R_{2,2}) &\mapsto d_{V,O,\gamma}(R_{1,1}, R_{2,2}) \ such \ that \\
d_{V,O,\gamma}(R_{1,1}, R_{2,2}) &= \sqrt{\gamma d_{1,O}^2(R_{1,1}, R_{2,2}) + (1-\gamma)d_{2,O}^2(R_{1,1}, R_{2,2})}
\end{aligned}
$$

*is a metric for all $\gamma \in [0,1)$ in a differentiable manifold $V$. The parameter $\gamma$ is a weight of each distance.*

**Remark 4.1.** *When analyzing the caterpillar data (see results in Section 7), the distance in the manifold generated by the generalist-specialist and horizontal shift corresponded to the equal weights case (i.e. $\gamma = \frac{1}{2}$)*

$$
d_{V,O}(R_{1,1}, R_{2,2}) = \sqrt{\frac{1}{2} d_{1,O}^2(R_{1,1}, R_{2,2}) + \frac{1}{2} d_{2,O}^2(R_{1,1}, R_{2,2})}
$$

*We chose equal weights because there was no a-priori reason why one path should be weighted more than the other path. We repeated the decomposition by changing the distance from full weight on one distance ($\gamma = 0$) to full weight on the other distance ($\gamma = 1$) and the results of the decomposition were similar.*

### 4.3.4. Decomposition

We have finally defined all the tools for our proposed decomposition. Let $\tilde{R}_O = R_{\tilde{O}_1, \tilde{O}_2}$ be an estimate of a Fréchet mean (i.e. in the Fréchet mean set) then we can see how to decompose the variation with the distance $d_{V,O}$

$$
\begin{aligned}
\widetilde{SSM}_O &= \sum_{i=1}^{n} d_{V,O}^2(R_{i,i}, \tilde{R}_O) \\
d_{V,O}^2(R_{i,i}, \tilde{R}_O) &= \frac{1}{2} d_{1,O}^2(R_{i,i}, \tilde{R}_O) + \frac{1}{2} d_{2,O}^2(R_{i,i}, \tilde{R}_O) \\
d_{1,O}^2(R_{i,i}, \tilde{R}_O) &= \mathrm{Arcd}^2(R_{i,0}, R_{\tilde{O}_1,0}) + \mathrm{Arcd}^2(R_{0,i}, R_{0,\tilde{O}_2}) \\
d_{2,0}^2(R_{i,i}, \tilde{R}_O) &= \Big( \mathrm{sign}(\alpha_i - \alpha_0)\mathrm{Arcd}(R_{i,i}, R_{0,i}) \\
&\quad - \mathrm{sign}(\tilde{O}_1 - \alpha_0)\mathrm{Arcd}(\tilde{R}_O, R_{0,\tilde{O}_2}) \Big)^2 \\
&\quad + \Big( \mathrm{sign}(\beta_i - \beta_0)\mathrm{Arcd}(R_{i,i}, R_{i,0}) \\
&\quad - \mathrm{sign}(\tilde{O}_2 - \beta_0)\mathrm{Arcd}(\tilde{R}_O, R_{\tilde{O}_1,0}) \Big)^2
\end{aligned}
$$

The advantage of defining the distance $d_{V,O}$ by $d_{1,O}$ and $d_{2,O}$ is that we can reorganize the terms of $\widetilde{SSM}_O$ such that

$$\widehat{SSM}_O = \widehat{SSM}_{1,O} + \widehat{SSM}_{2,O}$$

where

$$
\begin{aligned}
\widehat{SSM}_{1,O} = {} & \frac{1}{2}\sum_{i=1}^n \mathrm{Arcd}^2(R_{i,0}, R_{\tilde{O}_1,0}) + \frac{1}{2}\sum_{i=1}^n \Big( \mathrm{sign}(\alpha_i - \alpha_0)\mathrm{Arcd}(R_{i,i}, R_{0,i}) \\
& - \mathrm{sign}(\tilde{O}_1 - \alpha_0)\mathrm{Arcd}(\tilde{R}_O, R_{0,\tilde{O}_2}) \Big)^2
\end{aligned}
$$

and

$$
\begin{aligned}
\widehat{SSM}_{2,O} = {} & \frac{1}{2}\mathrm{Arcd}^2(R_{0,i}, R_{0,\tilde{O}_2}) + \frac{1}{2}\Big( \mathrm{sign}(\beta_i - \beta_0)\mathrm{Arcd}(R_{i,i}, R_{i,0}) \\
& - \mathrm{sign}(\tilde{O}_2 - \beta_0)\mathrm{Arcd}(\tilde{R}_O, R_{\tilde{O}_1,0}) \Big)^2
\end{aligned}
$$

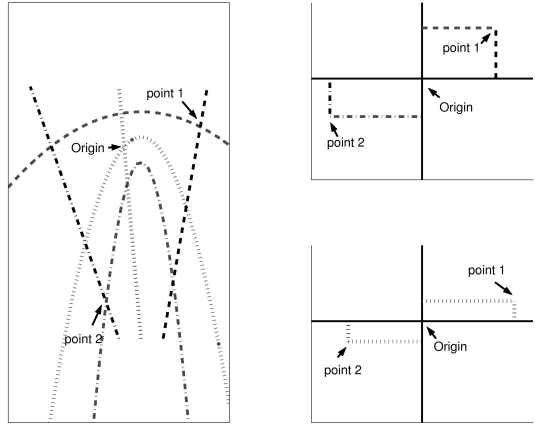

FIG 6. *Example of a 2 dimensional space of variation in the general case and representation of the corresponding two transformations. Left: Representation of the space of variation when the two modes are the horizontal shift and the generalist-specialist. The three points in the space of variation correspond to three different location and width parameters. For illustration, one of the points is taken to be the origin in the manifold. Top right: Representation of the second mapping of the space of variation into $\mathbb{R}^2$. This transformation preserves the arcdistances of curves which cross at the points of interest. Bottom right: Representation of the second mapping of the space of variation into $\mathbb{R}^2$. This transformation preserves the arcdistances of curves crossing at the origin.*

For a given origin $O$, the term $\widehat{SSM}_{1,O}$ quantifies the variation along the mode parameterized by $\alpha$ and the term $\widehat{SSM}_{2,O}$ quantifies the variation along the mode parameterized by $\beta$.

### 4.4. Multiple simultaneous modes, general case

We constructed in Subsection 4.2 a metric and a decomposition for multiple dimensional case satisfying the linearly separable case. More generally, when have $d'$ simultaneous modes of variation, we might have the linearly separable property satisfied for a subset of parameters but not for all parameters. For example, in model (1.1), we can write $R((w, m, h), t) = R_1((w, m), t) + R_2(h, t)$ where $R_1((w, m), t) = w \times z(w(t - m))$ and $R_2(h, t) = h$. We can define a distance in this three dimensional manifold as

$$d_V(R((w_1, m_1, h_1), \mathbf{t}), R((w_2, m_2, h_2), \mathbf{t}))$$

$$= \sqrt{d_V^2(R_1((w_1, m_1), \mathbf{t}), R_1((w_2, m_2), \mathbf{t}) + ||h_1 - h_2||^2}$$

where $d_V(R_1((w_1, m_1), \mathbf{t}), R_1((w_2, m_2), \mathbf{t})$ is the metric defined for a two-dimensional space of variation in Subsection 4.3.

Using the same tools we developed in Subsection 4.3 we can generalize the above metric and decomposition to a $d'$ dimensional space where the equality of path property is not necessarily satisfied by any subset of parameters. In a 2-dimensional manifold, we found two possible mappings of $V$, each preserving one (out of two) possible 2-steps paths between two points along the curves of variation. In the general $d'$-dimensional manifold, the number of mappings is the number of possible $d'$-step paths between two points along the curves of variation which is $(d'!)$. For each possible path $i$, we can define a mapping $L_{i,O}$, which allows us to define a function $d_{i,O}$. We define the metric in the manifold to be $d_O$ with weights $\gamma_i$ associated to each path metric. More precisely, for any two points $R_1$ and $R_2$ in the manifold

$$d_O^2(R_1, R_2) = \frac{1}{d'!} \sum_{i=1}^{d'!} \gamma_i d_{i,O}^2(R_1, R_2); \sum_i \gamma_i = 1.$$

Each $d_{i,O}^2$ is the sum of $d'$ terms, each depending on one parameter. Thus, we can reorganize the sums of square in the previous formula to be the sum of $d'$ terms, each depending on one parameter.

## 5. Consistency and choice of origin

We state in this section the main theorems proving the consistency of our estimates with the distances defined in the previous section. We also discuss our choice of origin in the space of variation used in the non-separable case.

### *5.1. Consistency*

Theorem 5.1 states the uniqueness of the Fréchet mean in the one dimensional case and the $d'$-dimensional case linearly separable case. We use Theorem 2.3 from Bhattacharya and Patrangenaru (2003) to prove consistency of our estimates in the general case in Proposition 5.1.

**Theorem 5.1.** *For i.i.d random variables $R_1, \ldots, R_n$ of measure $\mu$ in $(V, d_V)$ of finite Fréchet mean and Fréchet variance. For $V$ a one dimensional differentiable* *manifold or a d′ dimensional differentiable manifold linearly separable. We have that the Fréchet mean $\tilde{R}_F$ is unique, for all n its estimate $\tilde{R}$ from the data is unique and*

$$d_V(\tilde{R}, \tilde{R}_F) \rightarrow 0 \ (a.s)$$

$$\left| \frac{1}{n} \widetilde{SSM} - \widetilde{Var}_F \right| \rightarrow 0 \ (a.s)$$

See proof in the Appendix. Showing consistency in the general case is not trivial. We use Theorem 2.3 from Bhattacharya and Patrangenaru (2003) which establishes consistency of the Fréchet mean estimates for a complete manifold, to prove the following corollary

**Proposition 5.1.** *For all fixed origins O in a metric manifold $(V, d_{V,O})$*

$$\frac{1}{n} \widetilde{SSM}_O \longrightarrow \widetilde{Var}_{F,O} a.s$$

### *5.2. Choice of an origin*

We propose to use an origin $\tilde{O} = R((\alpha_0, \beta_0), \mathbf{t})$ in a compact set $K$ in $V$ which is a minimizer of the Fréchet Variance, i.e.

$$\widetilde{\mathrm{Var}}_{F,\tilde{O}} = \min_{O \in K} \widetilde{\mathrm{Var}}_{F,O}, \text{ where } \widetilde{\mathrm{Var}}_{F,O} = \int d_{V,O}^2(R, \tilde{R}_{F,O}) d\mu(R),$$

and $\tilde{R}_{F,O}$ is an element of the Fréchet mean set. This choice of origin is conservative, it will guarantee that the estimate of the variance in the manifold will not be inflated. A question of interest is, can this variance be estimated from the data when this origin is not set in advance? Let $F_n(O)$ be the estimate of the Fréchet variance from the data, i.e.

$$F_n(O) = \frac{1}{n} \widetilde{SSM}_{F,O} = \frac{1}{n} \sum_{i=1}^{n} d_O^2(R_{i,i}, \tilde{R}_{n,O})$$

where $\tilde{R}_{n,O}$ is the estimate of the Fréchet mean associated to the origin $O$ and the data $R_{1,1}, \ldots, R_{n,n}$. An estimate of $F_n(O)$ from the sample is the minimum of $F_n(O)$ over a compact set, i.e. the estimate is

$$F_n(O_n) = min_{O \in K} F_n(O)$$

The following proposition states that the sequence $F_n(O_n)$ converges to the desired variance function $\widetilde{\mathrm{Var}}_{F,\tilde{O}}$.

**Proposition 5.2.** *The Fréchet sample estimate at the sample minimizing origin $O_n$ converges to the Fréchet variance at the minimizing origin $\tilde{O}$, i.e. $F_n(O_n) \rightarrow \widetilde{Var}_{F,\tilde{O}}$.*

See proof in Appendix.

## 6. Implementation and extensions

We have discussed in this paper a useful decomposition of the variation in the manifold of variation. In the motivated model, the data does not lie in the manifold but is projected onto the manifold. More specifically, the data $z_i \in \mathbb{R}^d$ is such that $z_i = R_i + \epsilon_i$ where $R_i$ is the projection of $z_i$ onto the manifold and $\epsilon_i$ is additive error. The projection is the point in the manifold which is closest to the data in $\mathbb{R}^d$, i.e. which minimizes $SSE$ such that

$$||z_i - R_i|| = \min_{x \in V} ||z_i - x||$$

This projection might not be unique but we can use diagnostic plots shown in Fig. 7 to detect multiplicity of fits. We can also use local information to choose between multiple projections, so that for data which are close together in $\mathbb{R}^d$, their projections onto $V$ should be close together. We also suppose in this paper that the common shape function is known or could be estimated from the data. For the caterpillar example, the common shape function is estimated by a polynomial in an iterative procedure. Each iteration includes two steps, in each step parameters are fit to minimize $SSE$. Given initial parameters of variation, the first step finds the optimal values of the parameters of the polynomial. These polynomial coefficients are then used in the second step to find the optimal values of the parameters of variation. The parameters found in this iteration are used as initial parameters of the following iteration and this procedure is repeated until the algorithm converges to an optimal solution. This iterative procedure could be easily generalized for a common shape in any parametric family satisfying the condition of Theorem 2.1. This iterative procedure could also be generalized to estimate non-parameterically the common shape as well as the parameters of variation using algorithms described in Wang and Gasser (1997, 1999).

Finding the geodesic distance along one-dimensional manifolds is central to our method. This distance could be computed analytically using Equation 3.2. Another approach based on approximation is to consider the path length between two points in a one-dimensional manifold as the sum of lengths of small segments along the curve. The length of each small segment is approximated by the corresponding Euclidean distances.

## 7. Results

Our TMV analysis has been implemented and used by biologists to decompose variation along modes of interest in thermal performance curves of caterpillars in Izem and Kingsolver (2005), and viruses in Knies et al. (2006). The fitted parameters and fitted template polynomial optimized the weighted sums of squared error, weighted by the sample sizes of each family. The results of the decomposition on the motivating caterpillar data are presented in Table 1. Fig. 7 shows a visual diagnostic of the fits.

The data have only six measurements per curve, as shown in Fig. 1. Thus, we assumed for simplicity that the template shape is a polynomial of degree 4 in the results shown in Table 1 and the top two panels of Fig. 7. For identifiability, we assumed further that the fitted height parameters sum to zero. The polynomial fit to each curve as well as the template shape diagnostic fit are shown in the top two panels of Fig. 7. We see in the top left panel of Fig. 7 that the variation of the fitted polynomials closely matches the variation in the original data shown in Fig. 1. The top right panel of Fig. 7 shows the warped data compared to the fitted template shape polynomial of degree 4. Each curve $i$ was warped by its fitted parameters $(\hat{w}_i, \hat{m}_i, \hat{h}_i)$ so that all curves could be compared on the same scale to the fitted template shape. More precisely, a warped curve $i$ represents the warped growth rate $(\mathbf{z}_i - \hat{h}_i)/\hat{w}_i$ plotted at the warped temperature $(\mathbf{t} - \hat{m}_i)\hat{w}_i$, where $\mathbf{z}_i$ and $\mathbf{t}$ represent the observed growth rate and temperature vectors of curve $i$. We found that this representation is an excellent graphic diagnostic of the model fit. Under the assumptions of our model, after warping, we expect that the global maxima of all curves is at 0, that the warped curves are aligned and that the template shape is lying in the middle of the warped curves. We see in the top right panel of Fig. 7 that these expectations are mostly met. One of the curves appear to be an outlier, it corresponds to a family with small sample size. Although the fitted polynomial lies in the middle of the warped curves, the template curve seem to slightly overestimate the growth rate at low temperatures, which is due to the limitation of an oversmooth polynomial fit. In contrast to this good fit, the two bottom panels of Fig. 7 show the warped data compared to two fitted template shapes of a polynomial of degree 6 in the left panel and of a polynomial of degree 3 in the right panel. These fits are not as good as the polynomial of degree 4 fit because we see that in both panels the fitted maxima (at zero for warped data) is outside of the measurement range for many curves and the warped curves are not aligned. Because the curves are not aligned, a small set of curves contribute to fitting a large range of the polynomial of degree 6 (colder temperatures) and different groups of curves fit different regions of the template polynomial of degree 3.

We see in Table 1 that our model explains about 5/6 of the variation in the data, which is surprisingly good since all the modes have biological meaning. Very little variation is occurring along the vertical shift which suggests that selection of caterpillars with high growth rate at a particular temperature in one generation, will not result in individuals with high growth rate overall temperatures in other generations. Our decomposition separates the contribution of the generalist-specialist variation (27.11%) from the contribution of the horizontal shift variation (45.76%), which was not possible using Euclidean methods such as PCA and conventional ANOVA. We tested the robustness of our results using 500 bootstrap samples, where each bootstrap sample was obtained by sampling with replacement the family curves. We fitted the parameters and derived the decomposition for each bootstrap sample and we see in Table 1 that our decomposition of the variation hold.

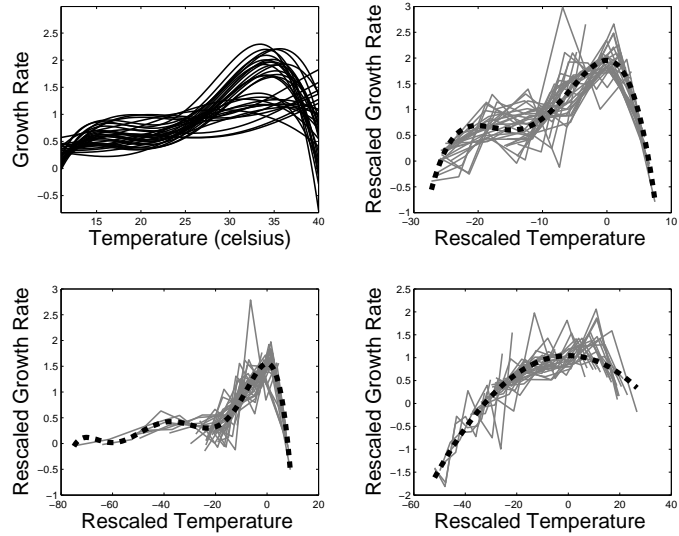

FIG 7. *Top left: Fitted curves to caterpillar data. Fitted curves are transformations of a common template by model 1.1. Top right: Fitted template shape of degree 4, $z(t) = -4.92 \times 10^{-5} - 2.41 \times 10^{-3} \times t - 0.032 \times t^2 + 2.19 \times t^4$, and warped data. The template lies in the middle of the rescaled data, it fits the data fairly well. Bottom left: Fitted template shape of degree 6 and warped data. Bottom right: Fitted template shape of degree 3 and warped data.*

TABLE 1
*Decomposition of the Variation and Bootstrap Intervals. Percentages of variation are computed according to Equation 5.1*

|  | vertical shift | horizontal shift | gen-spec | Total-Model |
|---|---|---|---|---|
| **Result** | **11.24** | **45.76** | **27.11** | **84.11** |
| Bootstrap (mean,sd) | 10.41,5.46 | 41.71,7.48 | 29.16,8.01 | 81.28,7.74 |
| Bootstrap (median, 5th,95th percentiles) | 10.64 2.87,19.88 | 27.86 27.86,52.81 | 29.57 11.8,40.60 | 82.89 67.47,83.38 |

## Acknowledgements

The research of the authors was supported in part by the National Science Foundation DMS-0308331, IBN-0212798, and EF-0328594 and in part by the National Institute of Health NIH/NLM LM07071.

## Appendix A: Appendix

**Theorem 2.1.** *For a fixed sampling vector $t \in \mathbb{R}^d$, let the function $R(.,\mathbf{t})$ be such that*

$$R(.,\mathbf{t}) : \mathbb{R}^{d'} \rightarrow V$$
$$\theta \mapsto R(\theta,\mathbf{t})$$

*If $R(.,\mathbf{t})$ is an homeomorphism from $\mathbb{R}^{d'}$ to $V$, then $V$ is a manifold of dimension $d'$.*

*Proof.* To show that $V$ is a manifold of dimension $d'$ we need to show that it is a separable topological space and that every neighborhood $\Omega$ in $V$ is homeomorphic to a neighborhood in $\mathbb{R}^{d'}$. $V$ is a topological subspace of $\mathbb{R}^d$ with topology induced by the topology in $(\mathbb{R}^d, ||.||)$. Since $\mathbb{R}^{d'}$ is separable, and $R(.,\mathbf{t})$ is an homeomorphism then $V$ is also separable. Since $R(.,\mathbf{t})$ is an homeomorphism from $(\mathbb{R}^{d'}, ||.||)$ to $(V, ||.||)$, then every neighborhoods $\Omega$ in $V$ is homeomorphic to a neighborhood in $\mathbb{R}^{d'}$. $\qquad\blacksquare$

**Theorem 4.1.** *For a $d'$ $(d' \geq 2)$ differentiable manifold of variation $V$ generated by a linearly separable model the non-negative function defined as*

$$d_V : V \times V \quad \rightarrow \quad \mathbb{R}^+$$
$$(R((\theta_{1,1}, \ldots, \theta_{d',1}), \mathbf{t}), R((\theta_{1,2}, \ldots, \theta_{d',2}), \mathbf{t})) \quad \mapsto \quad \sqrt{\sum_{i=1}^{d'} C_{\theta_{i,1}, \theta_{i,2}}^2}$$

*is a distance in $V$*

*Proof.* We will show this result in the two dimensional case. In dimension $d'$, the result would follow by induction using the same inequalities as for the two dimensional case. In two dimensions, $d_V$ is defined as

$$d_V : V \times V \quad \rightarrow \quad \mathbb{R}^+$$
$$(R((\alpha_1, \beta_1), \mathbf{t}), R((\alpha_2, \beta_2), \mathbf{t})) \quad \mapsto \quad \sqrt{C_{\alpha_1, \alpha_2}^2 + C_{\beta_1, \beta_2}^2}$$

where

$$C_{\alpha_1, \alpha_2} = \operatorname{Arcd}(R((\alpha_1, \beta), \mathbf{t}), R((\alpha_2, \beta), \mathbf{t})) \,\forall \beta, \text{ and}$$
$$C_{\beta_1, \beta_2} = \operatorname{Arcd}(R((\alpha, \beta_1), \mathbf{t}), R((\alpha, \beta_2), \mathbf{t})) \,\forall \alpha$$

We need to show that *(i)* $d_V$ is non-negative and symmetric, *(ii)* $d_V(x, y) = 0$ iff $x = y$, and *(iii)* $d_V$ satisfies the triangular inequality. Property *(i)* is satisfied by definition. Let us show property *(ii)*: if $d_V(R((\alpha_1, \beta_1), \mathbf{t}), R((\alpha_2, \beta_2), \mathbf{t})) = 0$, then

$$C_{\alpha_1, \alpha_2} = C_{\beta_1, \beta_2} = 0, \text{ So, } .$$

$$\operatorname{Arcd}(R((\alpha_1, \beta), \mathbf{t}), R((\alpha_2, \beta), \mathbf{t})) = 0, \text{ and}$$
$$\operatorname{Arcd}(R((\alpha, \beta_1), \mathbf{t}), R((\alpha, \beta_2), \mathbf{t})) = 0, \forall \alpha, \forall \beta$$

Since Arcd is a distance, this implies that $R((\alpha_1, \beta), \mathbf{t}) = R((\alpha_2, \beta), \mathbf{t})$ for all $\beta$, and $R((\alpha, \beta_1), \mathbf{t}) = R((\alpha, \beta_2), \mathbf{t})$ for all $\alpha$. Since $R$ is an homeomorphism, these two equalities imply that $\alpha_1 = \alpha_2$, and $\beta_1 = \beta_2$. So, $R((\alpha_1, \beta_1), \mathbf{t}) = R((\alpha_2, \beta_2), \mathbf{t})$. Let us show property *(iii)*, the triangular inequality. Let $R_i = R((\alpha_i, \beta_i), \mathbf{t})$, we can show that

$$d_V(R_1, R_3) \leq d_V(R_1, R_2) + d_V(R_2, R_3)$$

is equivalent to

$$d_V^2(R_1, R_3) \leq d_V^2(R_1, R_2) + d_V^2(R_2, R_3) + 2 \times d_V(R_1, R_2)d_V(R_2, R_3). \quad (A.1)$$

We have that,

$$
\begin{aligned}
\text{L.H.S of (A.1)} &= C_{\alpha_1,\alpha_3}^2 + C_{\beta_1,\beta_3}^2 \text{ (by definition)} \\
&\leq C_{\alpha_1,\alpha_2}^2 + C_{\alpha_2,\alpha_3}^2 + 2 \times C_{\alpha_1,\alpha_2}C_{\alpha_2,\alpha_3} \\
&\quad + C_{\beta_1,\beta_2}^2 + C_{\beta_2,\beta_3}^2 + 2 \times C_{\beta_1,\beta_2}C_{\beta_2,\beta_3} \text{ (b/c Arcd is a metric)} \\
&\leq d_V^2(R_1, R_2) + d_V^2(R_2, R_3) + 2 \times C_{\alpha_1,\alpha_2}C_{\alpha_2,\alpha_3} + 2 \times C_{\beta_1,\beta_2}C_{\beta_2,\beta_3}
\end{aligned}
$$

Then, to show (A.1), it is sufficient to show that

$$C_{\alpha_1,\alpha_2}C_{\alpha_2,\alpha_3} + C_{\beta_1,\beta_2}C_{\beta_2,\beta_3} \leq d_V(R_1, R_2)d_V(R_2, R_3)$$

which is equivalent to

$$
\begin{aligned}
&C_{\alpha_1,\alpha_2}^2 C_{\alpha_2,\alpha_3}^2 + C_{\beta_1,\beta_2}^2 C_{\beta_2,\beta_3}^2 + 2 \times C_{\alpha_1,\alpha_2}C_{\alpha_2,\alpha_3}C_{\beta_1,\beta_2}C_{\beta_2,\beta_3} \\
&\leq d_V^2(R_1, R_2)d_V^2(R_2, R_3)
\end{aligned}
$$

which is equivalent to

$$2 \times C_{\alpha_1,\alpha_2}C_{\alpha_2,\alpha_3}C_{\beta_1,\beta_2}C_{\beta_2,\beta_3} \leq C_{\alpha_1,\alpha_2}^2 C_{\beta_2,\beta_3}^2 + C_{\alpha_2,\alpha_3}^2 C_{\beta_1,\beta_2}^2$$

which is equivalent to

$$0 \leq (C_{\alpha_1,\alpha_2}C_{\beta_2,\beta_3} - C_{\alpha_2,\alpha_3}C_{\beta_1,\beta_2})^2$$

Since the last inequality is always true, $d_V$ satisfies the triangular inequality. From (i), (ii), and (iii) we have that $d_V$ is a metric in $V$. $\qquad\square$

**Proposition 4.2.** *When $V$ is linearly separable, $d_V$ is the distance defined in Subsection 4.2.2, and $d_{1,O}$ and $d_{2,O}$ are as defined in Section 4, we have that $d_{1,O} = d_{2,O} = d_V$ for all origins $O$ in the manifold.*

*Proof.* We first show the equality for $d_{1,O}$, then for $d_{2,O}$. When $V$ is linearly separable, we have by Corollary 4.1 that

$$d_{1,O}^2(R_{1,1}, R_{2,2}) = Arcd^2(R_{1,O}, R_{2,O}) + Arcd^2(R_{O,1}, R_{O,2})$$

By definition of the linearly separable space, the first term $Arcd^2(R_{1,O}, R_{2,O})$ is equal to $C_{\alpha_1,\alpha_2}^2$ using the notation of Subsection 4.2 and the distance does not depend on the origin $O$. Similarly, the second term $Arcd^2(R_{O,1}, R_{O,2})$ is equal to $C_{\beta_1,\beta_2}^2$ using the notation of Subsection 4.2 and the distance does not depend on the origin $O$. Thus, $d_{1,O}$ is equal to the distance $d_V$ defined in Theorem 4.1. The formula for $d_{2,O}(R_{1,1}, R_{2,2})$ given in Subsection 4.3.2 simplifies in the linearly separable case to

$$
\begin{aligned}
d_{2,O}(R_{1,1}, R_{2,2}) &= \sqrt{A^2 + B^2}, \text{ where} \\
A &= \text{sign}(\alpha_1 - \alpha_0)C_{\alpha_1,\alpha_0} - \text{sign}(\alpha_2 - \alpha_0)C_{\alpha_2,\alpha_0} \\
&= C_{\alpha_1,\alpha_2}; \text{ and} \\
B &= \text{sign}(\beta_1 - \beta_0)C_{\beta_1,\beta_0} - \text{sign}(\beta_2 - \beta_0)C_{\beta_2,\beta_0} \\
&= C_{\beta_1,\beta_2};
\end{aligned}
$$

Thus, $d_{2,O}$ is equal to the distance $d_V$ defined in Theorem 4.1. $\qquad\square$

**Proposition 4.4.** *The non-negative function $d_{V,O,\gamma}$ defined as*

$$
\begin{aligned}
d_{V,O,\gamma} : V \times V &\rightarrow \mathbb{R}^+ \\
(R_{1,1}, R_{2,2}) &\mapsto d_{V,O,\gamma}(R_{1,1}, R_{2,2}) \\
\text{such that} \\
d_{V,O,\gamma}(R_{1,1}, R_{2,2}) &= \sqrt{\gamma d_{1,O}^2(R_{1,1}, R_{2,2}) + (1-\gamma)d_{2,O}^2(R_{1,1}, R_{2,2})}
\end{aligned}
$$

*is a metric for all $\gamma \in [0,1)$ in a differentiable manifold $V$*

*Proof.* We prove this proposition in the general case that $d$ is a distance where $d = \sqrt{\gamma d_1^2 + (1-\gamma)d_2^2}$, $d_1$ is a distance and $d_2$ satisfies properties $(i)$ to $(iii)$ in Proposition 4.2. We need to show that $d$ is $(i)$ positive and symmetric, *(ii)* $d(x,y) = 0$ iff $x = y$, and *(iii)* $d$ satisfies the triangular inequality. (i) is satisfied by definition. Let us show *(ii)*, if $d(x,y) = 0$, then $d_1(x,y) = 0$ and $d_2(x,y) = 0$ which implies that $x = y$ since $d_1$ is a distance. Let us show *(iii)*, the triangular inequality. We need to prove that

$$d(x,y) \leq d(x,z) + d(z,y)$$

or equivalently that

$$d^2(x,y) \leq d^2(x,z) + d^2(z,y) + 2d(x,z)d(z,y) \tag{A.2}$$

Let us consider the left hand side of the inequality

$$
\begin{aligned}
d^2(x,y) &= \gamma d_1^2(x,y) + (1-\gamma)d_2^2(x,y) \\
&\leq \gamma d_1^2(x,z) + \gamma d_1^2(z,y) \\
&\quad + 2\gamma d_1(x,z)d_1(z,y) + (1-\gamma)d_2^2(x,z) \\
&\quad + (1-\gamma)d_2^2(z,y) + 2(1-\gamma)d_2(x,z)d_2(z,y) \\
d^2(x,y) &\leq d^2(x,z) + d^2(z,y) \\
&\quad + 2\gamma d_1(x,z)d_1(z,y) + 2(1-\gamma)d_2(x,z)d_2(z,y)
\end{aligned}
$$

To prove the inequality (A.2), it is sufficient to prove that

$$\gamma d_1(x,z)d_1(z,y) + (1-\gamma)d_2(x,z)d_2(z,y) \leq d(x,z)d(z,y) \tag{A.3}$$

We will reason by equivalence, inequality (A.2) is equivalent to

$$(\gamma d_1(x,z)d_1(z,y) + (1-\gamma)d_2(x,z)d_2(z,y))^2 \leq d^2(x,z)d^2(z,y) \qquad \text{(A.4)}$$

On one hand,

$$
\begin{aligned}
\text{L.H.S of inequality (A.4)} \quad = \quad & \gamma^2 d_1^2(x,z)d_1^2(z,y) + (1-\gamma)^2 d_2^2(x,z)d_2^2(z,y) \\
& + 2\gamma(1-\gamma)d_1(x,z)d_1(z,y)d_2(x,z)d_2(z,y)
\end{aligned}
$$

On the other hand,

$$
\begin{aligned}
\text{R.H.S of inequality (A.4)} \quad = \quad & \left(\gamma d_1^2(x,z) + (1-\gamma)d_2^2(x,z)\right) \\
& \times \left(\gamma d_1^2(z,y) + (1-\gamma)d_2^2(z,y)\right) \\
= \quad & \gamma^2 d_1^2(x,z)d_1^2(z,y) + (1-\gamma)^2 d_2^2(x,z)d_2^2(z,y) \\
& + \gamma(1-\gamma)\left(d_1^2(x,z)d_2^2(z,y) + d_1^2(z,y)d_2^2(x,z)\right)
\end{aligned}
$$

After canceling the common terms from the left and the right hand side of inequality (A.4), we have the inequality

$$0 \leq \gamma(1-\gamma)\left(d_1(x,z)d_2(z,y) + d_1(x,z)d_2(z,y)\right)^2$$

Since this last inequality is always true, then the triangular inequality is always true. Since the function $d$ satisfies *(i), (ii)*, and *(iii)* then it is a metric in $M$. ◻

**Theorem 5.1.** *For i.i.d random variables $R_1, \ldots, R_n$ of measure $\mu$ in $(V, d_V)$ of finite Fréchet mean and Fréchet variance, and $V$ one dimensional differentiable manifold or $d'$ dimensional differentiable linearly separable manifold. We have that the Fréchet mean $\tilde{R}_F$ is unique, its estimate from the data $\tilde{R}$ is unique for all $n$ and*

$$
\begin{aligned}
d_V(\tilde{R}, \tilde{R}_F) \quad &\rightarrow \quad 0 \ \text{(a.s)} \\
\left|\frac{1}{n}\widetilde{SSM} - \widetilde{Var_F}\right| \quad &\rightarrow \quad 0 \ \text{(a.s)}
\end{aligned}
$$

*Proof.* To show the uniqueness and convergence in the one dimensional case, we first map the manifold into $\mathbb{R}$ using an isometry $L$ and then we use known consistency of the mean results in $\mathbb{R}$ to prove consistency in $M$. Let $\mathcal{L}$ be the class of functions, $\mathcal{L} = \{L_\theta, \theta \in \Theta\}$ such that for all $\theta_0$ in $\Theta$

$$
\begin{aligned}
L_{\theta_0} : V \quad &\rightarrow \quad R \\
x = R(\theta, \mathbf{t}) \quad &\mapsto \quad L_{\theta_0}(x) \\
\text{such that } L_{\theta_0}(x) \quad &= \quad \text{sign}(\theta - \theta_0)\text{Arcd}(R(\theta,\mathbf{t}), R(\theta_0,\mathbf{t}))
\end{aligned}
$$

For a given parameter $\theta_0$, we can consider $R(\theta_0, \mathbf{t})$ as the *origin* in the manifold $V$ by the transformation $L_{\theta_0}$ because by definition $L_{\theta_0}(R(\theta_0,\mathbf{t})) = 0$. For all $\theta_0$, $L_{\theta_0}$ satisfies the isometry property, i.e.

$$|L_{\theta_0}(R_1) - L_{\theta_0}(R_2)| = \text{Arcd}(R_1, R_2),$$

and the distance between two points $L_{\theta_0}(R_1)$ and $L_{\theta_0}(R_2)$ does not depend on the origin. By definition of the space of variation and by construction of the transforming function $L_\theta$, this function is a homeomorphism. i.e.

(i) Continuity of $L_{\theta_0}$: If $\mathrm{Arcd}(x_n, x)$ converges to 0, then by isometry $|L_{\theta_0}(x_n) - L(x)|$ converges to 0. So, $L_{\theta_0}$ is continuous.

(ii) Invertibility of $L_{\theta_0}$: We need to show that $L_{\theta_0}(x_1) = L_{\theta_0}(x_2) \Rightarrow x_1 = x_2$. By isometry the left hand side implies that $\mathrm{Arcd}(x_1, x_2) = 0$, which implies that $x_1 = x_2$ (because Arcd is a metric in $V$).

(ii) Continuity of the inverse of $L_{\theta_0}$: We need to show that

$$|L_{\theta_0}(x_n) - L_{\theta_0}(x)| \to 0 \Rightarrow \mathrm{Arcd}(x_n, x) \to 0.$$

This property follows directly from the isometry.

Moreover, $L$ is such that $L_{\theta_1}(x) = L_{\theta_0}(x) + \mathrm{sign}(\theta_1 - \theta_0)L_{\theta_1}(\theta_0)$ Using $L$, let us show that the Fréchet mean is unique. We have that

$$E(L_{\theta_0}(R)) = \mathrm{Argmin}_{x \in \mathbb{R}} \int (L_{\theta_0}(R) - x)^2 d\mu$$

Let $R_0 = L_{\theta_0}^{-1}(E(L_{\theta_0}(R_i)))$, we can show that $R_0$ is the Fréchet mean. Let $F(R)$ be the Fréchet function, then

$$\forall R \neq R_0, F(R) \quad > \quad \int (L_{\theta_0}(R_i) - E(L_{\theta_0}(R)))^2 d\mu, \text{ and by isometry,}$$

$$> \quad \int \mathrm{Arcd}^2(R_i, R_0) d\mu$$

So, $R_O$ is the only element in the Fréchet mean set. We can repeat the same argument to show uniqueness of the Fréchet sample mean $\tilde{R}$ and the empirical measure.

Since by the law of large numbers,

$$\overline{L_{\theta_0}} \to E(L_{\theta_0}(R))(a.s) \tag{A.5}$$

then by isometry of $L_{\theta_0}$, we have that

$$d_V(\tilde{R}, \tilde{R}_F) \to 0 \text{ (a.s)}$$

Similarly, since $(L_{\theta_0}(R_i) - E(L_{\theta_0}(R_i)))^2$ are i.i.d and of finite mean $Var_F$, then by law of large numbers

$$\left| \frac{1}{n} \sum (L_{\theta_0}(R_i) - E(L_{\theta_0}(R_i)))^2 - \mathrm{Var}_F \right| \to 0 \text{ (a.s)} \tag{A.6}$$

On the other hand,

$$\frac{1}{n}\widetilde{SSM} - \mathrm{Var}_F = \left( \frac{1}{n} \sum (L_{\theta_0}(R_i) - E(L_{\theta_0}(R_i)))^2 - \mathrm{Var}_F \right) + (\overline{L_{\theta_0}} - E(L_{\theta_0}(R)))^2 \tag{A.7}$$

From equality (A.7), and using the two convergence results in (A.6) and (A.5), we have that

$$\left| \frac{1}{n} \widetilde{SSM} - \mathrm{Var}_F \right| \to 0 \text{ (a.s)}$$

In the $d'$ linearly separable case, as in the one-dimensional case, we define a mappings $L_{(\theta_{1,0}, \ldots, \theta_{d',0})} \in \mathcal{L}$ (associated with origins parameterized by $(\theta_{1,0}, \ldots, \theta_{d',0})$), which transform the nonlinear space to a Euclidean space. i.e.

$$
\begin{aligned}
L_{(\theta_{1,0}, \ldots, \theta_{d',0})} : V &\rightarrow \mathbb{R}^{d'} \\
x = R((\theta_1, \ldots, \theta_{d'}), \mathbf{t}) &\mapsto L_{(\theta_{1,0}, \ldots, \theta_{d',0})}(x) \\
L_{(\theta_{1,0}, \ldots, \theta_{d',0})}(x) &= \Big( \mathrm{sign}(\theta_{1,0} - \theta_1) \times C_{\theta_1, \theta_{1,0}}, \ldots, \\
&\qquad \mathrm{sign}(\theta_{d',0} - \theta_{d'}) \times C_{\theta_{d',0}, \theta_{d'}} \Big)
\end{aligned}
$$

For all $(\theta_{1,0}, \ldots, \theta_{d',0})$, $L_{(\theta_{1,0}, \ldots, \theta_{d',0})}$ is an homeomorphism from $(V, d_V)$ to $\mathbb{R}^{d'}$. The proof in this case is equivalent to the proof in the one-dimensional case. Since $d_V$ is a metric and $L_{(\theta_{1,0}, \ldots, \theta_{d',0})}$ satisfies the isometry property, then we can show that it is continuous, invertible and the inverse is continuous. Similarly, the proof of uniqueness of Fréchet mean estimate and Fréchet mean is equivalent to the one dimensional case. $\qquad \square$

**Corollary 5.1.** *For all fixed origins $O$ in a metric manifold $(V, d_{V,O})$*

$$\frac{1}{n} \widetilde{SSM}_O \longrightarrow \widetilde{Var}_{F,O} a.s$$

*Proof.* Let $F_n(x)$ be the estimate of the Fréchet function from the sample, i.e.

$$F_n(x) = \frac{1}{n} \sum_{i=1}^{n} d_V^2(R_i, x)$$

Let $x_n$ be a minimizer of $F_n$, i.e.

$$F_n(x_n) = \min_{x \in V} F_n(x)$$

To use Theorem 2.3 of (Bhattacharya & Patrangenaru (2003)), we need to show that $(V, d_V)$ is a complete metric space and all bounded closed set is compact.

1. Completeness: Let $x_n$, and $x_m$ two points in $V$, then by definition of $d_V$

$$d_V(x_n, x_m) = \sqrt{\gamma d_1^2(x_n, x_m) + (1 - \gamma) d_2^2(x_n, x_m)}$$

So, if $(x_n)$ is a cauchy sequence in $(V, d_V)$, it is cauchy in $(V, d_1)$ and $(V, d_2)$. We know by construction of $d_1$ and $d_2$, $(V, d_1)$ and $(V, d_2)$ are both homeomorphic to $(\mathbb{R}^2, ||.||)$ (where $||.||$ denotes the norm derived from the Euclidean metric). By this homeomorphism, a cauchy sequence $(x_n)$ in $(V, d_1)$ converges to $a_1$ in $V$. Similarly, a cauchy sequence $(x_n)$ in $(V, d_2)$ converges to $a_2$ in $V$. Let us finally show that $a_1 = a_2$. Since $V$ is a subspace of $\mathbb{R}^{d'}$, we have that

$$||x_n - a_1|| \leq d_1(x_n, a_1) \tag{A.8}$$

(b/c the Euclidean distance is the shortest distance), so $(x_n)$ converges to $a_1$ in $\mathbb{R}^{d'}$. Similarly, we have that

$$||x_n - a_2|| \leq d_2(x_n, a_2) \tag{A.9}$$

(b/c the Euclidean distance is the shortest distance), so $(x_n)$ converges to $a_2$ in $\mathbb{R}^{d'}$. By equations (A.8) and (A.9), and using the triangular inequality of the Euclidean distance, we have that

$$||a_1 - a_2|| = 0$$

So, $a_1 = a_2$. Then, $d_V(x_n, a_1)$ converges, so $(x_n)$ converges.

2. Compact sets in $(V, d_V)$. Similarly, using the homeomorphism, a closed and bounded set $A$ in $(V, d_V)$ is closed and bounded in $(V, d_1)$ and $(V, d_2)$. By the homeomorphisms to $(\mathbb{R}^2, ||.||)$, the space $A$ is compact in $(V, d_1)$ and $(V, d_2)$, so it is compact in $(V, d_V)$.

So, using Theorem 2.3 of (Bhattacharya and Patrangenaru (2003)), we have that

$$(x_n) \to R_F (a.s)$$

where $R_F$ is a Fréchet mean. On the other hand, we have that

$$F_n(R_F) \to Var_F$$

Since

$$|F_n(x_n) - Var_F| \leq |F_n(x_n) - F_n(R_F)| + |F_n(R_F) - Var_F|$$

We have that

$$F_n(x_n) \to \mathrm{Var}_F (a.s) \qquad \square$$

**Proposition 5.1.** *The Fréchet sample estimate at the sample minimizing origin $O_n$ converges to the Fréchet variance at the minimizing origin $\tilde{O}$, i.e. $F_n(O_n) \to \widetilde{Var}_{F,\tilde{O}}$.*

*Proof.* Let $F(O) = \widetilde{\mathrm{Var}}_{F,O}$. Then, by the property of minimums $F_n(O_n)$ and $F(\tilde{O})$ and continuity of $F_n$ and $F$ we have that,

First, $\forall \epsilon, \exists (\alpha, \beta)$ s.t.

$$F_n(\alpha) - \epsilon/2 < \quad F_n(O_n), \qquad \text{and}$$
$$F(\beta) - \epsilon/2 < \quad f(\tilde{O}).$$

Second,

$$F_n(O_n) \quad \leq \quad F_n(\beta) \text{ because } O_n \text{ is a minimizer of } F_n, \text{ and}$$
$$F(\tilde{O}) \quad \leq \quad F(\alpha) \text{ because } \tilde{O} \text{ is a minimizer of } F.$$

In addition, from Proposition 5.1, we know that $F_n$ converges to $F$ a.s. Thus,

$$\exists N \text{ s.t. } \forall n \geq N, \quad F_n(\alpha) - F(\alpha) \quad > -\epsilon/2, \text{ and}$$
$$F_n(\beta) - F(\beta) \quad < \epsilon/2.$$

Finally, by combining all the inequalities we have that

$$\forall \epsilon, \exists N, \text{ s.t. } \forall n \geq N, -\epsilon < F_n(O_n) - F(\tilde{O}) < \epsilon.$$

which proves the convergence. □

## References

Bhattacharya, R. and V. Patrangenaru (2003). Large sample theory of intrinsic and extrinsic sample means on manifolds. I. *Ann. Statist. 31*(1), 1–29. MR1962498

Bhattacharya, R. and V. Patrangenaru (2005). Large sample theory of intrinsic and extrinsic sample means on manifolds. II. *Ann. Statist. 33*(3), 1225–1259. MR2195634

Brumback, L. C. and M. J. Lindstrom (2004). Self modeling with flexible, random time transformations. *Biometrics 60*(2), 461–470. MR2066281

Castro, P. E., W. H. Lawton, and E. A. Sylvestre (1986). Principal modes of variation for processes with continuous sample curves. *Technometrics 28*, 329–337.

de Silva, V. and J. B. Tenenbaum (2003a). Local versus global methods for nonlinear dimensionality reduction. In S. Becker, S. Thrun, and K. Obermayer (Eds.), *Advances in Neural Information Processing Systems*, Volume 15, pp. 705–712. Cambridge: MIT Press.

de Silva, V. and J. B. Tenenbaum (2003b). Unsupervised learning of curved manifolds. In *Nonlinear estimation and classification (Berkeley, CA, 2001)*, Volume 171 of *Lecture Notes in Statist.*, pp. 453–465. New York: Springer. MR2005810

Donoho, D. L. and C. Grimes (2003). Hessian eigenmaps: locally linear embedding techniques for high-dimensional data. *Proc. Natl. Acad. Sci. USA 100*(10), 5591–5596 (electronic). MR1981019

Dryden, I. L. and K. V. Mardia (1998). *Statistical shape analysis*. Wiley Series in Probability and Statistics: Probability and Statistics. Chichester: John Wiley & Sons Ltd. MR1646114

Fletcher, P., C. Lu, S. Pizer, and S. Joshi (2004). Principal geodesic analysis for the study of nonlinear statistics of shape. *IEEE transactions on medical imaging 23*(8), 995–1005.

Fréchet, M. (1948). Les éléments aléatoires de nature quelconque dans un espace distancié. *Ann. Inst. H. Poincaré 10*, 215–310. MR0027464

Härdle, W. and J. S. Marron (1990). Semiparametric comparison of regression curves. *The Annals of Statistics 18*, 63–89. MR1041386

Hastie, T. and W. Stuetzle (1989). Principal curves. *J. Amer. Statist. Assoc. 84*(406), 502–516. MR1010339

Izem, R., , and J. Marron (2003). Quantifying nonlinear modes of variation of curves. In *Proceedings of the International Statistical Institute 54th session [CD-ROM], August 13-20. Berlin, Germany.* Invited talk.

Izem, R. (2004, May). *Analysis of Nonlinear Variation in Functional Data.* Ph.D. thesis, Dept. Statistics and Operations Research, University of North Carolina at Chapel Hill, Chapel Hill, NC.

Izem, R. and J. Kingsolver (2005). Variation in continuous reaction norms: quantifying directions of biological interest. *The American Naturalist 166*(8), 277–289.

Kingsolver, J., R. Gomulkiewicz, and P. Carter (2001). Variation, selection and evolution of function-valued traits. *Genetica 112-113*, 87–104.

Kingsolver, J., G. Ragland, and J. Schlichta (2004). Quantitative genetics of continuous reaction norms: Thermal sensitivity of caterpillar growth rates. *Evolution 58*(7), 1521–1529.

Kneip, A. and J. Engel (1995). Model estimation in nonlinear regression under shape invariance. *The Annals of Statistics 23*, 551–570. MR1332581

Kneip, A. and T. Gasser (1988). Convergence and consistency results for self-modeling nonlinear regression. *The Annals of Statistics 16*, 82–112. MR0924858

Knies, J., R. Izem, K. Supler, J. Kingsolver, and B. C. (2006). Continuous reaction norm analysis of temperature adaptation in bacteriophage. *PLoS Biology 4*(7), 1257–1264.

Kume, A. and H. Le (2000). Estimating Fréchet means in Bookstein's shape space. *Adv. in Appl. Probab. 32*(3), 663–674. MR1788088

Ladd, W. M. and M. J. Lindstrom (2000). Self-modeling for two-dimensional response curves. *Biometrics 56*(1), 89–97.

Lawton, W. H., E. A. Sylvestre, and M. S. Maggio (1972). Self modeling nonlinear regression. *Technometrics 14*, 513–532.

Le, H. (2001). Locating Fréchet means with application to shape spaces. *Adv. in Appl. Probab. 33*(2), 324–338. MR1842295

Lindstrom, M. J. (1995). Self-modelling with random shift and scale parameters and a free-knot spline shape function. *Statistics in Medicine 14*, 2009–2021.

Ramsay, J. O. and B. W. Silverman (2002). *Functional data analysis: methods and case studies.* Springer Series in Statistics. Springer. MR1910407

Ramsay, J. O. and B. W. Silverman (2005). *Functional data analysis, 2nd edition* (Second ed.). Springer Series in Statistics. New York: Springer. MR2168993

Scheiner, S. (1993). Genetics and evolution of phenotypic plasticity. *Annual Review of Ecology and Systematics 24*, 35–68.

Silverman, B. W. (1995). Incorporating parametric effects into functional principal components analysis. *J. Roy. Statist. Soc. Ser. B 57*(4), 673–689. MR1354074

Tenenbaum, J. B., V. d. Silva, and J. C. Langford (2000). A Global Geometric Framework for Nonlinear Dimensionality Reduction. *Science 290*(5500), 2319–2323.

Wang, K. and T. Gasser (1997). Alignment of curves by dynamic time warping. *Ann. Statist. 25*(3), 1251–1276. MR1447750

Wang, K. and T. Gasser (1999). Synchronizing sample curves nonparametrically. *Ann. Statist. 27*(2), 439–460. MR1714722